\documentclass[aps,prl,floatfix,twocolumn,reprint]{revtex4}
\usepackage{eurosym}
\usepackage{amsbsy}
\usepackage{latexsym,epsfig,graphicx}
\usepackage{dcolumn}
\usepackage{graphicx}
\usepackage{subfigure}
\usepackage{comment}
\usepackage{color}
\usepackage{bm}
\usepackage{mathrsfs}
\usepackage{amssymb}
\usepackage{amsfonts}
\usepackage{mathrsfs}
\usepackage{amsmath}
\usepackage{color}
\usepackage{graphicx}
\usepackage{bm}
\usepackage{amssymb}
\usepackage{xspace}
\usepackage{epstopdf}
\usepackage{dcolumn}
\usepackage{tabularx}
\usepackage{longtable}
\usepackage[colorlinks=true, letterpaper=true, pdfstartview=FitV, linkcolor=blue, citecolor=blue, urlcolor=blue]{hyperref}
\usepackage[normalem]{ulem}
\usepackage[version=4]{mhchem}

\begin{document}

\title{Majorana Doublets, Flat Bands, and Dirac Nodes in \emph{s}-Wave
Superfluids}
\author{Haiping Hu}
\affiliation{Department of Physics, The University of Texas at Dallas, Richardson, Texas
75080, USA}
\author{Fan Zhang}
\affiliation{Department of Physics, The University of Texas at Dallas, Richardson, Texas
75080, USA}
\author{Chuanwei Zhang}
\email{chuanwei.zhang@utdallas.edu}
\affiliation{Department of Physics, The University of Texas at Dallas, Richardson, Texas
75080, USA}

\begin{abstract}
Topological superfluids protected by mirror and time-reversal symmetries are exotic
states of matter possessing Majorana Kramers pairs (MKPs), yet their
realizations have long been hindered by the requirement of unconventional
pairing. We propose to realize such a topological superfluid by utilizing $s$%
-wave pairing and emergent mirror and time-reversal symmetries in two
coupled 1D ultracold atomic Fermi gases with spin-orbit coupling. By
stacking such systems into 2D, we discover topological and Dirac-nodal
superfluids hosting distinct MKP flat bands. We show that 
the emergent symmetries make the MKPs and their flat bands stable against
pairing fluctuations that otherwise annihilate paired Majoranas. Exploiting
new experimental developments, our scheme provides a unique platform for
exploring MKPs and their applications in quantum computation.
\end{abstract}

\maketitle

{\color{blue}{\em Introduction.}}---Spin-orbit coupling (SOC) plays a
crucial role in many topological quantum phenomena of condensed matter
physics~\cite{Kane2010RMP,Qi2011RMP}. In ultracold atomic gases, SOC has
been experimentally realized by coupling different hyperfine ground states
through counter-propagating Raman lasers~\cite%
{1dsoc1,1dsoc2,1dsoc3,1dsoc4,1dsoc5,1dsoc6,1dsoc7,1dsoc8,soc1,soc2,soc3}.
Due to their highly controllability and free of disorder, the spin-orbit
coupled ultracold atomic gases have opened a broad avenue for exploring
novel topological quantum matter. In particular, the cooperation of three
key ingredients, i.e., SOC, Zeeman coupling, and $s$-wave pairing
interaction, can produce effective $p$-wave superfluids~\cite%
{coldpro1,coldpro2,coldpro3} that host Majorana excitations~\cite%
{majorana1,majorana2,majorana3}. Because of their non-Abelian braiding
statistics and potential applications in fault-tolerant quantum computing~%
\cite{topoquantum}, topological defects containing unpaired Majoranas have
been extensively studied in solid-state systems nowadays~\cite%
{pro1,pro2,pro3,pro4,pro5,pro6,pro7,pro8,pro9,pro10,exp1,exp2,exp3,exp4,exp5,exp6,socclass}.

These superfluids with unpaired Majoranas belong to class D 
in the ten-fold way of Altland-Zirnbauer classification~\cite%
{class1,class2}. Without additional symmetries, the coupling between two
Majoranas can lift their zero-energy degeneracy. Time-reversal (TR) symmetry 
($\mathcal{T}^{2}=-1$) can, however, dictate them to form
a Kramers doublet, dubbed Majorana Kramers pair (MKP)~\cite%
{tritscqxl,tritsc1,tritsc2,tritsc3}. Topological superfluids hosting protected MKPs
belong to a completely distinct symmetry class, i.e., the DIII or mirror class~\cite{tritsc2}.
Intriguingly, MKPs enjoy symmetry-protected non-Abelian braiding
statistics~\cite{FZ-CK,xiongjunliuprx}, which may constitute advantages for
quantum computing. 

There have been several tantalizing proposals for realizing topological superconductors 
hosting MKPs in solid-state materials~\cite%
{tritscqxl,tritsc1,tritsc2,tritsc3,FZ-CK,xiongjunliuprx,F1,F2,tritsc5,tritsc6,tritsc7,tritsc8,tritsc9,tritsc10,tritsc11,tritsc12,tritsc13,tritsc15,tritsc16}%
, such as those proximitized devices exploiting the {unconventional} $s_{\pm
}$-wave~\cite{tritsc1,tritsc16}, $d_{x^{2}-y^{2}}$-wave~\cite{tritsc3}, or spatially sign-switching pairing~\cite{pro1}. 
However, these schemes are challenging, as they
strongly rely on the presence of exotic pairing and its fine control in
materials~\cite{no-go}. In this context, ultracold atomic gases may provide
a more controllable platform for exploring topological
superfluids hosting MKPs~\cite{tritsc2}. In contrast to extrinsic proximity-induced
superconductivity in solid-state platforms, superfluid orders in
ultracold atomic gases are formed through intrinsic $s$-wave attractive
interactions. In particular, a superfluid phase may be destroyed by quantum
fluctuations in a 1D chain, therefore it is crucial to exploit  
weakly-coupled 1D chains or 2D/3D arrays to suppress quantum fluctuations. 
Yet, it has been shown that couplings between identical class D (and even class BDI~\cite{socclass}) 
chains induce edge pairing fluctuations that destroy Majorana modes~\cite{wucongjun,Wu2}. 
Thus, two questions naturally arise. Can TR-invariant topological mirror superfluids be realized in
ultracold atomic gases with conventional $s$-wave pairing? 
If so, can TR and mirror symmetries protect MKPs from pairing fluctuations?
In this Letter, we address these two important questions by showing that the
remarkable physics of TR-invariant topological mirror superfluids and
associated MKPs can be realized in ultracold atomic gases by utilizing 
experimentally accessible $s$-wave pairing and synthetic 1D SOC 
\cite{1dsoc1,1dsoc2,1dsoc3,1dsoc4,1dsoc5,1dsoc6,1dsoc7,1dsoc8}. 
Here are our main findings.

\begin{figure}[t]
\centering
\includegraphics[width=3.3in]{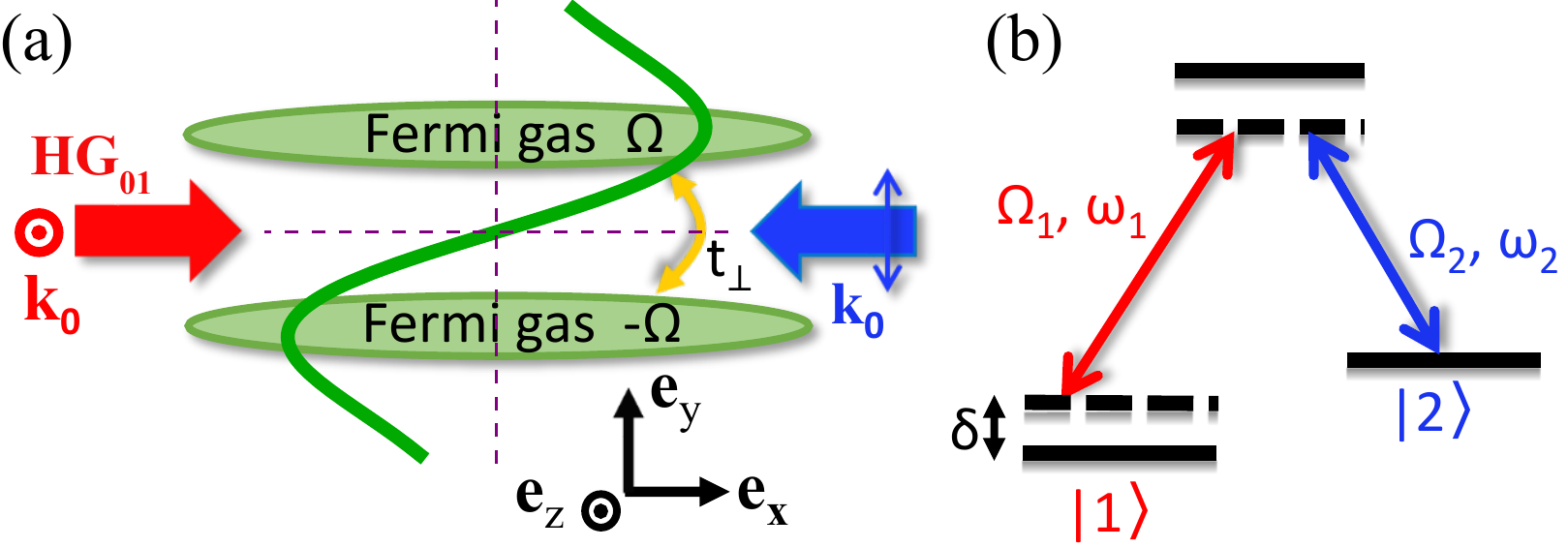}
\caption{Schematics of proposed experimental setups. (a) 1D SOC generated by
two counter-propagating Raman lasers along $\bm{e_{x}}$, i.e., one HG$_{01}$
beam (red arrow) polarized along $\bm{e_{z}}$ with frequency $\protect\omega%
_{1}$ and one Gaussian beam (blue arrow) polarized along $\bm{e_{y}}$ with
frequency $\protect\omega _{2}$. The green line shows the resulting Zeeman
field along $\bm{e_y}$. (b) Two-photon process induced by the two Raman
lasers in (a) with a detuning $\delta$.}
\label{tsc_model}
\end{figure}

First, although the Zeeman field from Raman coupling in synthetic SOC breaks
TR symmetry in a Fermi gas, effective TR and mirror symmetries emerge for two coupled gases with opposite Zeeman fields (Fig.~\ref{tsc_model}), 
which can be realized by changing the beam profile of one
Raman laser from Gaussian to Hermite-Gaussian~\cite{HG}. The emergent TR and
mirror symmetries, together with \textit{s}-wave pairing, can be exploited
to realize TR-invariant topological mirror superfluids~\cite{tritsc2}.

Second, by tuning the Zeeman field strength and chemical potential, our 1D
system undergoes various phase transitions between different phases and the
topological superfluid characterized by a $\mathbb{Z}_{2}$ invariant and
the emergence of MKPs. Even though the SOC is 1D, our 2D system exhibits
both topological and Dirac-nodal~\cite{DSC} superfluids hosting distinct
flat bands of MKPs. This extension strongly suppresses 
quantum fluctuations that may destroy the two superfluid phases.

Thirdly, as evidenced by our self-consistent calculations \cite{selfbdg6,selfbdg1,selfbdg2,selfbdg3,selfbdg4,selfbdg5}, the degeneracies
of MKPs and their flat bands are symmetry protected against pairing
fluctuations, which are known to annihilate paired Majoranas for coupled
1D chains. (All these results also apply to the 3D case.)
Therefore, our scheme provides a simple experimentally feasible route for realizing
TR-invariant topological and Dirac-nodal superfluids, paving the way
for observing MKPs and exploring their non-Abelian statistics~\cite{FZ-CK,xiongjunliuprx} 
and interaction effects~\cite{F1,F2}.

{\color{blue}{\em Model.}}---Consider two coupled 1D Fermi gases of
ultracold atoms with the same SOC but opposite Zeeman fields. (A double-well
trapping potential along ${\hat{y}}$ is used to create this 
system.) As sketched in Fig.~\ref{tsc_model}, the SOC can be achieved by
two counter-propagating Raman lasers coupling two atomic hyperfine 
states $|1\rangle $ and $|2\rangle $. This setup is the same as those in
previous experiments~\cite%
{1dsoc1,1dsoc2,1dsoc3,1dsoc4,1dsoc5,1dsoc6,1dsoc7,1dsoc8,soc1,soc2,soc3},
except that one laser beam is changed from Gaussian to Hermite-Gaussian HG$%
_{01}$ mode~\cite{HG}, and can be described by the Hamiltonian 
$h_{k}={\hbar ^{2}k^{2}}/{2m}+\Omega \sigma _{z}+\delta \sigma _{y}+2\alpha
k\sigma _{y}$ in a rotated basis with $\left\vert 1,2\right\rangle =\left(
\left\vert \uparrow \right\rangle \pm i\left\vert \downarrow \right\rangle
\right) /\sqrt{2}$. Here $k$ is the quasi-momentum in each gas, 
$\alpha $ is the SOC strength, $\delta$ is the two-photon detuning,
and $\Omega ={\Omega }_{0}y\exp \left( -y^{2}/w^{2}\right) $ is the
position-dependent Raman coupling serving as the Zeeman field. Given the
antisymmetric HG$_{01}$ beam, the Zeeman field is opposite at the two gases,
which is crucial for realizing an emergent TR symmetry.

Taking into account the \emph{s}-wave interaction induced superfluidity, the
physics of our 1D Fermi gas system can be described by the Bogoliubov-de
Gennes (BdG) Hamiltonian \cite{SM} $H_{k}=\Psi_{k}^{\dag }\mathcal{H}_{k}^{BdG}\Psi_{k}/2$ with%
\begin{equation}
\mathcal{H}_{k}^{\mathrm{BdG}}=\big[\xi _{k}+2\alpha \sin k\,\sigma
_{y}-t_{\perp }s_{x}\big]\tau _{z}+\Omega \sigma _{z}s_{z}+\Delta \tau _{x}
\label{BdG}
\end{equation}%
expressed in the Nambu spinor basis $\Psi _{k}=(\phi _{k},i\sigma _{y}\phi
_{-k}^{\dag })$. Here $\phi _{k}\!=\!(c_{k\uparrow ,1},c_{k\downarrow
,1},c_{k\uparrow ,2},c_{k\downarrow ,2})^{T}$ with $c_{k\sigma ,s}$ the
fermion annihilation operators; ${\bm\sigma }$, ${\bm s}$, and ${\bm\tau }$
are Pauli matrices acting on the fermion spin, \emph{double chain}, and
particle-hole spaces, respectively; $\xi _{k}=-2t\cos k-\mu $ is the
intra-chain kinetic energy with a chemical potential $\mu $, $t_{\perp }$ is
the inter-chain coupling, and $\delta =0$ has been chosen for the detuning.
The lattice regularization of the free-space fermion kinetic energy would
not change any essential physics \cite{SM}. Importantly, the Zeeman field $%
\Omega \sigma _{z}s_{z}$ is exactly opposite for the two chains, and the 
\emph{s}-wave pairing order parameter $\Delta $ must be self-consistently
determined~\cite{selfbdg6,selfbdg1,selfbdg2,selfbdg3,selfbdg4,selfbdg5}.
Diagonalizing the Hamiltonian~(\ref{BdG}), we obtain the quasiparticle
energy spectrum 
\begin{eqnarray}
\!\!\!\!\!\!E(k) &=&\pm \bigg[(2\alpha \sin k\pm t_{\perp })^{2}+\Omega
^{2}+\Delta ^{2}+\xi _{k}^{2}  \notag \\
&&\pm 2\sqrt{(\Delta ^{2}+\xi _{k}^{2})\Omega ^{2}+(2\alpha \sin k\pm
t_{\perp })^{2}\xi _{k}^{2}}\bigg]^{1/2},  \label{QPE}
\end{eqnarray}%
with two-fold degeneracies at $k=0$ and $\pi $ due to an emergent TR
symmetry, as we elaborate below.

{\color{blue}{\em Symmetry and invariant.}}---The model~(\ref{BdG}) 
has three \emph{independent} symmetries that govern the
underlying physics. First, there is an intrinsic particle-hole symmetry
reflecting the BdG redundancy: $\mathcal{P}\mathcal{H}_{k}^{\mathrm{BdG}}%
\mathcal{P}^{-1}=-\mathcal{H}_{-k}^{\mathrm{BdG}}$ with $\mathcal{P}=\tau
_{y}\sigma _{y}\mathcal{K}$ and $\mathcal{K}$ the complex conjugation.
Second, even though the TR symmetry is explicitly broken by the Zeeman field
within each chain, Eq.~(\ref{BdG}) is still invariant under TR followed by
chain inversion, i.e.,%
\begin{equation}
\mathcal{\tilde{T}}\,\mathcal{H}_{k}^{\mathrm{BdG}}\,\mathcal{\tilde{T}}^{-1}=\mathcal{H}%
_{-k}^{\mathrm{BdG}}\,,~~~~\mathcal{\tilde{T}}=is_{x}\sigma _{y}\mathcal{K}.
\label{TRS}
\end{equation}%
Given that $\mathcal{\tilde{T}}^{2}=-1$, such an emergent TR symmetry dictates the
Kramers degeneracies found in the spectrum~(\ref{QPE}) at $k=0$ and $\pi $.
Note that the composite operation of $\mathcal{P}$ and $\mathcal{\tilde{T}}$ also
leads to a chiral symmetry: $\mathcal{C}\mathcal{H}_{k}^{\mathrm{BdG}}%
\mathcal{C}^{-1}=-\mathcal{H}_{k}^{\mathrm{BdG}}$ with $\mathcal{C}=\mathcal{%
P}\mathcal{\tilde{T}}$. Thirdly, the setup has a mirror symmetry such
that the two chains are the mirror images of each other, i.e.,%
\begin{equation}
\mathcal{M}\,\mathcal{H}_{k}^{\mathrm{BdG}}\,\mathcal{M}^{-1}=\mathcal{H}%
_{k}^{\mathrm{BdG}}\,,~~~~\mathcal{M}=is_{x}\sigma _{y}.  \label{MS}
\end{equation}%
Since the mirror symmetry with $\mathcal{M}^{2}=-1$ is a spatial symmetry,
naturally $[\mathcal{M},\mathcal{O}]=0$ with $\mathcal{O}=\mathcal{P}$,~$%
\mathcal{\tilde{T}}$~and~$\mathcal{C}$.

In light of the above symmetry analysis, the Hamiltonian~(\ref{BdG}) belongs
to both the DIII class~\cite{class1,class2} and the mirror class~\cite%
{tritsc2} in topological classification. It follows that a $\mathbb{Z}_{2}$
index $\nu$~\cite{z2invariant,SM} and a mirror winding number $\gamma_m$,
with $\nu ={\gamma_m}\mod 2$~\cite{tritsc2}, can both be used for
characterizing the band topology of model~(\ref{BdG}).

We find that the transitions between topologically distinct phases occur at
the phase boundary where 
\begin{equation}
\xi _{k}^{2}+\Delta ^{2}=\Omega ^{2},~~~4\alpha ^{2}\sin ^{2}k=t_{\perp
}^{2}.  \label{bc}
\end{equation}%
For $t_{\perp }=0$, the quasiparticle gap closes at $k=0$, and the phase
boundary reduces to that of single-chain superfluids~\cite{pro2,pro3}. 
For a finite $t_{\perp }$, the quasiparticle gap closes at a
finite $k$, and the critical Zeeman fields read
\begin{equation}
\Omega _{\pm }=[(2t\sqrt{1-t_{\perp }^{2}/4\alpha ^{2}}\pm \mu )^{2}+\Delta
^{2}]^{1/2}.  \label{CZeeman}
\end{equation}

Applying the established formulas for $\nu $~\cite{z2invariant,SM} and $%
\gamma _{m}$~\cite{tritsc2} to Eq.~(\ref{BdG}), we conclude that 
\begin{flalign}\label{index}
\nu=\gamma_m=\left\{
\begin{array}{cc}
                1 & ~~~\mbox{if}~~\Omega_-<|\Omega|<\Omega _+\,,\\
                0 & \mbox{otherwise}\,.\\
\end{array} \right.
\end{flalign}Our model in the nontrivial regime realizes not only
a TR-invariant topological superfluid but also the first
topological mirror superfluid~\cite{tritsc2} in degenerate gases.

{\color{blue}{\em Self-consistent phase diagram.}}---In ultracold atomic
gases, the local $s$-wave pair potential in real space must be determined in
a self-consistent manner~\cite{selfbdg6,selfbdg1,selfbdg2,selfbdg3,selfbdg4,selfbdg5}%
, together with the quasiparticle energies and wave functions. In our
numerical calculations \cite{SM}, the chemical potential is fixed without
loss of generality, and the open boundary condition is used for the purpose
of observing MKPs. We choose $L=120$ as the length of chain, $t$ as the
energy unit, and $\langle\Delta\rangle=\sum_{i}|\Delta _{i}|/L$ as the
pairing strength.

\begin{figure}[t]
\centering
\includegraphics[width=3.4in]{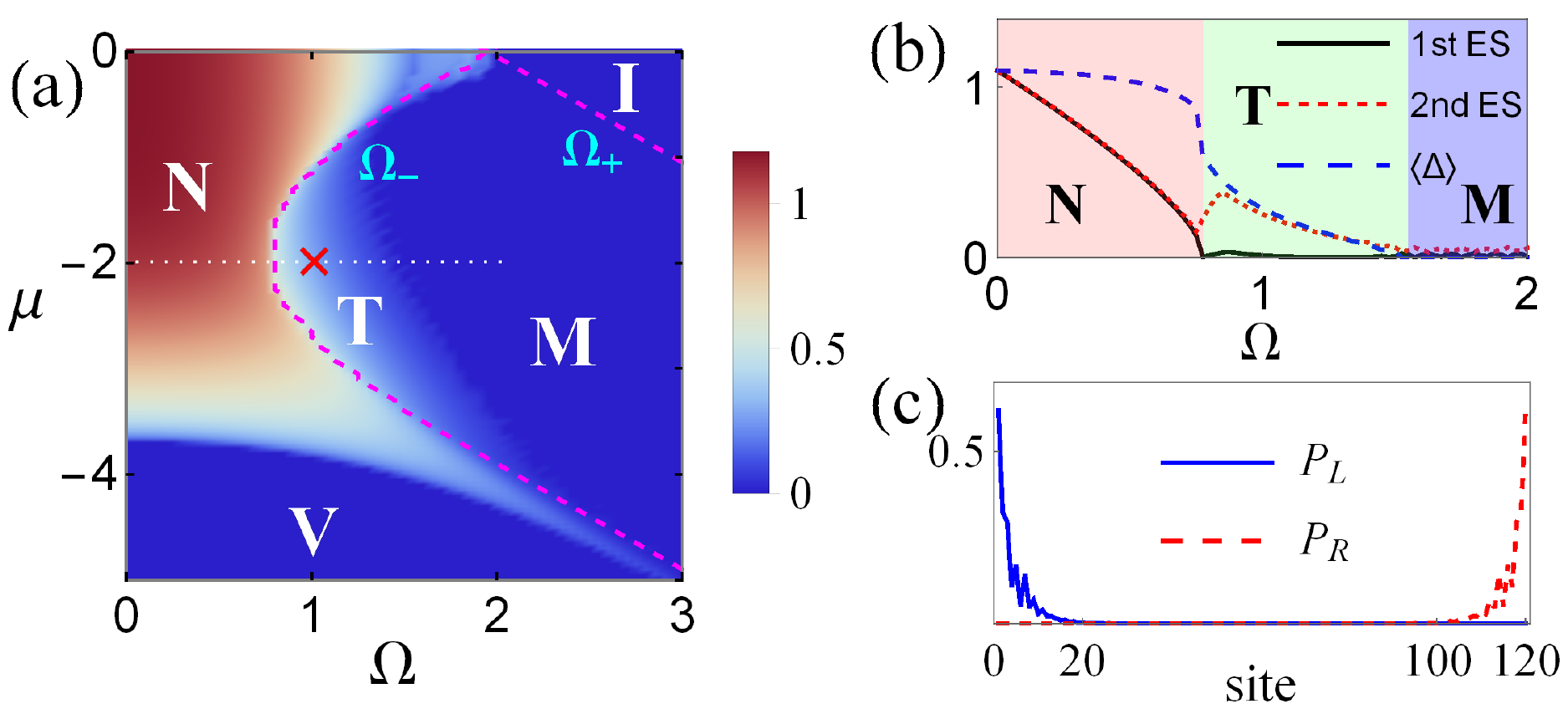}
\caption{(a) Phase diagram in the $\Omega$-$\protect\mu$ plane, symmetric
with respect to $\protect\mu =0$ and $\Omega=0$. The contour plot shows the
site-averaged pairing $\langle\Delta\rangle$ in the normal superfluid (N),
topological superfluid (T), metal with SOC (M), polarized insulator (I), and
trivial vacuum (V). The dotted red lines are the phase boundaries determined
by Eq.~(\protect\ref{bc}). (b) Phase transitions along the white dotted line
in (a). The black solid (red dotted) lines denote the first (second)
quasiparticle excitation states (ES) in the spectrum, both of which are
two-fold degenerate. (c) Probability distributions of the left (L) and right
(R) MKPs at the red cross in (a). $\sum_i P_L(i)=\sum_i P_R(i)=2$ are the
hallmarks of MKPs. $\protect\alpha =1$ and $t_{\perp}=0.5$ are used in
(a)-(c).}
\label{phasediag}
\end{figure}
Figure~\ref{phasediag}(a) plots the phase diagram in the $\Omega$-$\mu $
plane, which is symmetric with respect to $\mu=0$ and $\Omega=0$. Evidently,
the numerical phase boundaries are in good harmony with those determined by
Eq.~(\ref{bc}). In total, there are five distinct phases: the normal
superfluid, topological superfluid, metal with SOC, polarized insulator, and
trivial vacuum. The vacuum state occurs when $|\mu|$ is too large to cross
the single-particle bands. The system becomes the polarized insulator near $%
|\mu| =0$ if the Zeeman field strength $|\Omega|$ is sufficiently large;
each lattice site per chain is occupied by one fermion of the same
polarization. At relatively smaller $|\Omega|$ and $|\mu|$, superfluidity
spontaneously emerges with a finite bulk pairing gap for quasiparticle
excitations. In this regime, whereas it is the normal superfluid without any
boundary zero mode if both $|\Omega|$ and $|\mu|$ approach zero, it becomes
the topological superfluid with two degenerate zero modes per boundary,
i.e., the MKP, if $|\mu|$ approaches to the original band degeneracies and
if $|\Omega|>\Omega _-$ as required in Eq.~(\ref{index}). As $|\Omega|$
further increases, the superfluidity gradually vanishes, and the metal phase
emerges with an excitation gap scales linearly with $1/L$.

Figure~\ref{phasediag}(b) with $\mu=-2$ features the most appealing part of
the phase diagram, where there are two successive phase transitions as $%
\Omega$ increases from $0$. The first transition occurs at $\Omega=\Omega _-$%
: the normal superfluid turns to the topological superfluid with the
emergence of one localized MKP per boundary, as shown in Fig.~\ref{phasediag}%
(c). As $\Omega$ becomes stronger, the pairing strength $\langle\Delta%
\rangle $ becomes weaker. Eventually at the second transition, $%
\langle\Delta\rangle$ vanishes and the system enters into the metal phase
with gapless single-particle excitations.

{\color{blue}{\em 2D topological superfluids.}}---By stacking our
double chains, we can obtain exotic 2D and 3D topological
superfluids protected by the emergent TR and mirror symmetries. The
extension to higher dimensions can suppress quantum fluctuations 
and stabilize long-range pairing orders. We focus on the 2D case 
\cite{SM}, and the 3D generalization is straightforward. The staggered Zeeman field switches sign between neighboring
chains along $\hat{y}$. This setup can be described by the BdG Hamiltonian 
\begin{eqnarray}
\mathcal{H}_{\bm k}^{\mathrm{BdG}} &=&\big[\xi _{k_{x}}+2\alpha \sin
k_{x}\,\sigma _{y}-(t_{1}+t_{2}\cos k_{y})s_{x}  \notag \\
&&-t_{2}\sin k_{y}\,s_{y}\big]\tau _{z}+\Omega s_{z}\sigma _{z}+\Delta \tau
_{x},  \label{2dbdg}
\end{eqnarray}%
where $t_{1}$ and $t_{2}$ are the alternating inter-chain couplings along $%
\hat{y}$. Such a system has an emergent property 
\begin{equation}
\mathcal{\tilde{T}}\,\mathcal{H}^{\mathrm{BdG}}(k_{x},k_{y})\,\mathcal{\tilde{T}}^{-1}=%
\mathcal{H}^{\mathrm{BdG}}(-k_{x},k_{y}),\label{aT}
\end{equation}%
i.e., the system respects the TR symmetry in Eq.~(\ref{TRS}) and belongs to
class DIII with a $\mathbb{Z}_{2}$ invariant $\nu _{k_{y}}$ for any $k_{y}$,
which is an anomalous pumping parameter~\cite{FZ-CK}. 
\begin{figure}[t]
\centering
\includegraphics[width=3.1in]{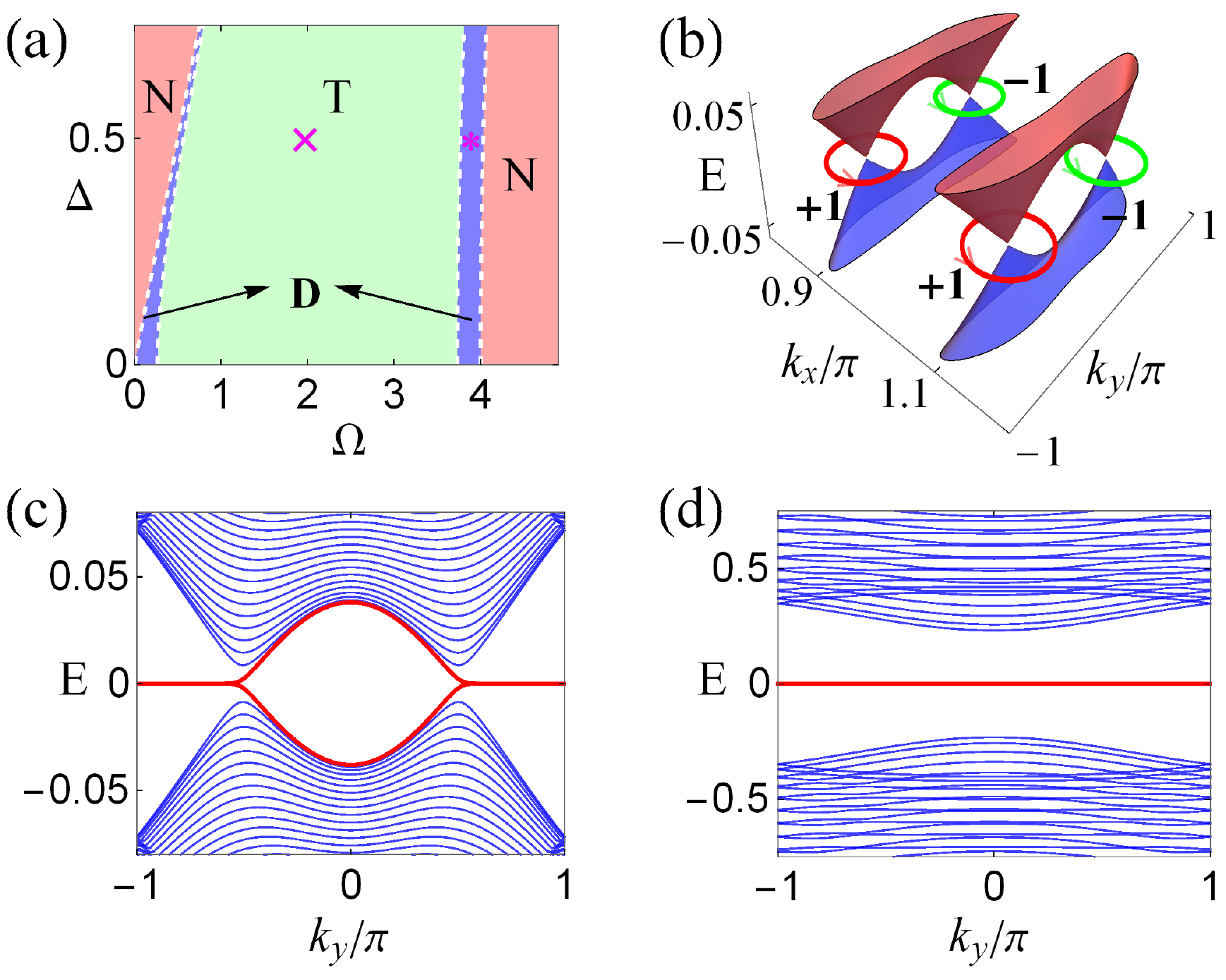}
\caption{(a) Phase diagram in the $\Omega $-$\Delta $ plane for the 2D
model~(\protect\ref{2dbdg}). The red, green, and blue regions denote the
normal (N), topological (T), and Dirac-nodal (D) superfluids, respectively.
(b) Bulk quasiparticle spectrum for the Dirac superfluid labeled by the red
star in (a). Each Dirac point is indexed by a winding number $\protect\gamma %
_{t}=\pm 1$. (c)-(d) Quasiparticle spectrum with MKP edge flat bands under open
boundary condition for the Dirac and topological superfluids labeled in (a). 
$t_{1}=t_{2}=0.5$, $\protect\alpha =1$, and $\protect\mu =-2$ are used in
(a)-(d).}
\label{fig3}
\end{figure}

Consequently, there can be three distinct phases for Eq.~(\ref{2dbdg}).
Whereas the superfluid is normal if $\nu_{k_y}=0$ for any $k_y$, an
unprecedented topological superfluid emerges if $\nu_{k_y}=1$ for any $k_y$.
Remarkably in the topological phase, there emerges a flat band of MKPs at
the edge along $\hat y$, because there is a MKP corresponding to the
nontrivial $\mathbb{Z}_{2}$ invariant for any $k_y$. (This edge flat band is a 
consequence of the bulk topological property, and the band flatness is protected 
by the TR and mirror symmetries, although the edge flat band itself may be trivial~\cite{flatband} 
if treated as a 1D system.) Intriguingly, if $%
\nu_0\neq\nu_{\pi}$, a nodal superfluid emerges. As the $\mathbb{Z}_{2}$
invariant changes from $k_y=0$ to $k_y=\pi$, the bulk gap must close at at
least one $k_y$ in between $0$ and $\pi$, separating the $\nu=0$ and $\nu=1$
regimes, and a flat band of MKPs emerge between the projected nodes~\cite{DSC} 
at the edge along $\hat y$. 

Figure~\ref{fig3}(a) illustrates a representative phase diagram in the $%
\Omega$-$\Delta$ plane. Indeed, all three phases emerge and the nodal
superfluid intervenes the normal and topological ones. Surprisingly, we find
that the nodes are Dirac points with linear dispersions and topological
protections. Diagonalizing Eq.~(\ref{2dbdg}) yields the phase boundaries and
the Dirac point positions, as determined by 
\begin{equation}
\xi_{k_{x}}^{2}+\Delta^{2}=\Omega^{2},~4\alpha^{2}%
\sin^{2}k_{x}=t_{1}^2+t_{2}^2+2t_1t_2\cos k_{y}.  \label{bc2}
\end{equation}
The Dirac points are two-fold degenerate and come in multiples of four, as
dictated by the $\mathcal{\tilde{T}}$ and $\mathcal{M}$ symmetries that respectively
flip $k_x$ and $k_y$. Moreover, any loop enclosing one such Dirac point has
a total winding number $\gamma_t=\pm 1$~\cite{DSC}, protected by an emergent
chiral symmetry 
\begin{eqnarray}
\widetilde{\mathcal{C}}\,\mathcal{H}^{\mathrm{BdG}}_{\bm k}\,\widetilde{%
\mathcal{C}}^{-1}=-\mathcal{H}^{\mathrm{BdG}}_{\bm k}\,, ~~~~\widetilde{%
\mathcal{C}}=\tau_y\sigma_y\,.  \label{CS}
\end{eqnarray}
Figure~\ref{fig3}(b) displays the four Dirac points and their $\gamma_t$'s
accordingly. Figures~\ref{fig3}(c) and~\ref{fig3}(d) contrast the MKP edge flat
bands in the Dirac-nodal and topological superfluids.

{\color{blue}{\em Discussion.}}---It is instructive to consider the
stability of MKPs and their flat bands in our proposed scheme. For
an array of topological superfluids without the $\mathcal{\tilde{T}}$ and $%
\mathcal{M}$ symmetries, it is known that Majoranas interactions spontaneously produce
nonuniform pairing fields $\Delta _{j}e^{i\phi _{j}}$ and edge supercurrent
loops~\cite{Wu2}. Since the phase fluctuations cannot be gauged away, the
Majoranas can be gapped out in pairs. Neglecting long-range interactions,
the Majorana annihilation is governed by the nearest-neighbor Josephson
couplings as follows~\cite{wucongjun}:%
\begin{equation}
\!\!\!\!\delta H\!=\!-\!\sum\nolimits_{\langle ij\rangle }[J_{0}\cos \phi
_{ij}+iJ_{ij}\gamma _{i}\gamma _{j}\sin (\phi _{ij}/2)],
\end{equation}%
with $J_{0},J_{ij}>0$ and $\phi _{ij}=\phi _{i}-\phi _{j}$. While the first
term favors a global phase coherence, the second term splits the
Majorana zero modes through phase fluctuations.

\begin{figure}[t]
\centering
\includegraphics[width=3.15in]{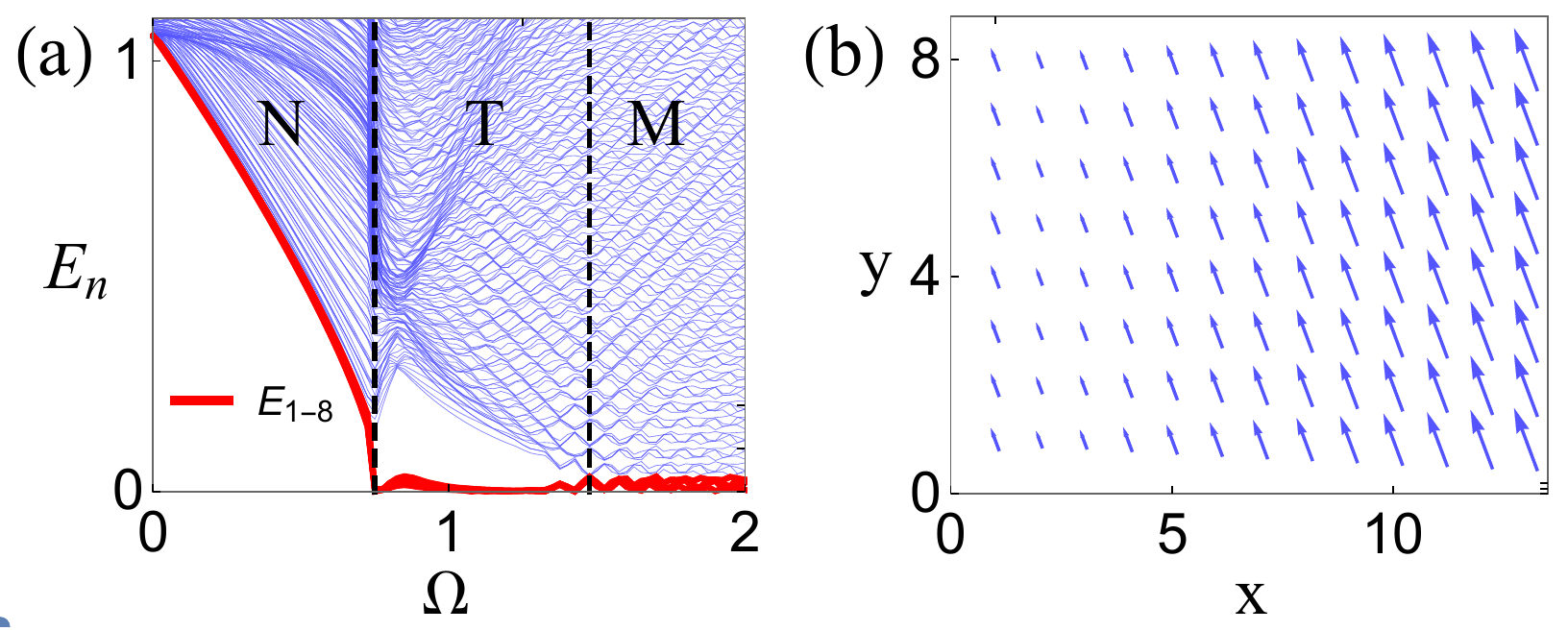}
\caption{(a) Self-consistent quasiparticle spectrum for the $100\times 8$
lattice model. The red lines denote the eight lowest quasiparticle
excitation states. (b) Vector plot of the local pairing fields $\Delta
_{j}e^{i\protect\phi _{j}}$ for $\Omega =1.1$. The length (direction) of
each arrow denotes the strength (phase) of the local pairing field. $%
t_{1}=t_{2}=0.5$, $\protect\alpha =1$, and $\protect\mu =-2$ are used in (a)-(b).}
\label{pairfield}
\end{figure}

In sharp contrast, the MKP flat bands of our system are robust against such
phase fluctuations. This can be best understood from the symmetry
perspective. Under the $\mathcal{M}$ operation, the local
pairing term $\Delta_i e^{i\phi_i}c_{i\uparrow}c_{i\downarrow}$ becomes $%
\Delta_i e^{i\phi_i}c_{i+1\uparrow}c_{i+1\downarrow}$ since the sublattice
and spin indices in Eq.~(\ref{2dbdg}) are simultaneously flipped. 
For the Josephson coupling, the $J_{ij}$-term must vanish as 
$\phi_i=\phi_{i+1}$ is dictated by mirror symmetry.

Our self-consistent calculations also agree with such a symmetry argument.
Fig.~\ref{pairfield}(a) plots the BdG spectrum for a $100\times 8$ lattice
model of Eq.~(\ref{2dbdg}). Consistent with Fig.~\ref{fig3}(a), the system
undergoes two transitions as the Zeeman field increases: from a normal
superfluid to a topological one and eventually to a metal phase with $%
\langle\Delta\rangle=0$. (Dirac points are absent due to the finite size
effect.) The topological phase hosts eight-fold degenerate zero modes on the
boundary along $\hat y$, forming a MKP flat band that is also stable against
the $t_1$-$t_2$ anisotropy. These remarkable features suggest 
that our proposed scheme is superior to previous ones. 

Finally, a few comments are in order on relevant experiments. In the 2D setup, the Zeeman
field switches sign between neighboring chains of distance $b$. This can be
realized through the periodic modulation $\Omega _{1}\sim \cos \left( \pi
y/b\right) $ for one Raman laser. Such a modulation can be produced by a
digital micromirror device~\cite{DM1,DM2,SM}, which can generate an
arbitrary modulation of laser intensity. This setup can be generalized to a
3D lattice with $\Omega _{1}\sim \cos \left( \pi y/b\right) \cos \left( \pi
z/c\right) $, where a boundary MKP flat band is anticipated. Our scheme of
restoring TR symmetry via a spatial reflection can be generalized to various
different systems, where the SOCs have been realized for other types of 
pseudospin states~\cite{Li2017,Li2016,Ye,Ye2}.

The MKPs can be experimentally detected using spatially resolved
radio-frequency spectroscopy~\cite{SM,rf1,rf2,rf3,ldosrf1,ldosrf2}, which
measures the local density of states, similar to scanning tunneling
microscope. Different from a single Majorana mode, 
the intrinsic two-fold degeneracy of a MKP can be further affirmed from the energy splitting 
and spatial separation of two Majoranas due to symmetry breaking~\cite{SM}, which can be
induced by the imbalance of $\Omega$ between the two chains. Our results not
only provide a simple experimental scheme for realizing mirror- and TR-invariant 
topological and Dirac-nodal superfluids but also establish a unique platform
for exploring MKPs and their applications in quantum computation.

\textit{Note added}.---{Near the submission of this manuscript, we became
aware of an independent work~\cite{tritsc14} that explores MKPs in double semiconductor nanowires 
with proximity-induced s-wave pairing and {\it ad hoc} opposite Zeeman fields. While pairing fluctuation,
mirror symmetry, Dirac phase, and flat band are not discussed in Ref.~\cite{tritsc14}, 
the results based on the emergent time-reversal symmetry in the two works agree with each other.

\begin{acknowledgments} 
H.H. and C.Z. are supported by NSF (PHY-1505496, PHY-1806227), ARO (W911NF-17-1-0128), AFOSR (FA9550-16-1-0387). 
F.Z. is supported by UTD (Research Enhancement Funds) and ARO (W911NF-18-1-0416).
\end{acknowledgments}

\clearpage
\onecolumngrid
\appendix
\section{Supplementary Materials}
\subsection{Topological Invariant and Mirror Symmetry}

In the main text, we have derived the $\mathbb{Z}_{2}$ topological invariant 
$\nu$ directly using the mirror symmetry $\mathcal{M}$. Here, we give more
details on the construction of this $\mathbb{Z}_2$ invariant, which
classifies time-reversal (TR) invariant topological superfluids in 1D and 2D.

\subsection{I. 1D Case}

The mirror symmetry $\mathcal{M}$ is a spatial symmetry with $\mathcal{M}%
^{2}=-1$. As $[\mathcal{M},\mathcal{H}_{k}^{BdG}]=0$, the Hamiltonian can be
decomposed into two sectors, with each sector belonging to a specific
subspace labeled by one of the two mirror eigenvalues $\pm i$.
Mathematically, the $8\times 8$ Hamiltonian Eq.~(1) in the main text can be
block-diagonalized as $U_{1}\mathcal{H}_{k}^{BdG}U_{1}^{\dag}=h(k)\oplus
h^{\ast }(-k)$ by the transformation $U_{1}$ such that $U_{1}\mathcal{M}%
U_{1}^{\dag }=i\,\mathrm{diag}(I_{4\times 4},-I_{4\times 4})$ and $h(k)$
reads 
\begin{equation}
h(k)=\left( 
\begin{array}{cccc}
-\xi _{k}+\Omega & -i\alpha _{k} & \Delta & 0 \\ 
i\alpha _{k} & -\xi _{k}-\Omega & 0 & \Delta \\ 
\Delta & 0 & \xi _{k}+\Omega & i\alpha _{k} \\ 
0 & \Delta & -i\alpha _{k} & \xi _{k}-\Omega%
\end{array}%
\right) ,
\end{equation}%
with $\xi _{k}=-2t\cos k-\mu $, $\alpha _{k}=2\alpha \sin k-t_{\perp }$. As
the eigenvalues $\pm i$ switch signs under the individual action of
anti-unitary TR or particle-hole operator, each eigen-block has neither TR
nor particle-hole symmetries, whereas it remains invariant under the
co-action of these two symmetries, i.e., the chiral symmetry $\mathcal{C}$.
Thus, each mirror eigen-block belongs to the AIII symmetry class in the
Altland-Zirnbauer table~\cite{tritsc2}, labeled by opposite 1D winding
numbers~\cite{tritsc2}. The latter is because the two eigen-blocks are
related by the TR symmetry. To see this fact, consider $|\phi\rangle$ as an
eigenstate of $\mathcal{M}$ with mirror eigenvalue $i$. Since $\mathcal{M}(\mathcal{\tilde{T}}|\phi \rangle )=\mathcal{\tilde{T}}\mathcal{M}|\phi \rangle =\mathcal{\tilde{T}}(i|\phi \rangle )=-i\mathcal{\tilde{T}}|\phi \rangle$, 
$\mathcal{\tilde{T}}|\phi \rangle $ is also an eigenstate of $\mathcal{M}$
but with mirror eigenvalue $-i$.

Now we focus on the $h(k)$ sector. The chiral symmetry operator can be
chosen as $\mathcal{C}_{\mathcal{M}=i}=\tau_y\otimes\sigma_y$ with $\{%
\mathcal{C}_{\mathcal{M}=i},\,h(k)\}=0$. To construct the topological
invariant, we can chose a unitary transformation $U_{2}$ such that $U_{2}%
\mathcal{C}_{\mathcal{M}=i} U_{2}^{\dag }=\mathrm{diag}(I_{2\times
2},-I_{2\times 2})$. In this new basis, $h(k)$ can be rewritten in the
off-diagonal form as follows: 
\begin{equation}
U_{2}h(k)U_{2}^{\dag }=\left( 
\begin{array}{cc}
0 & g(k) \\ 
g^{\dag }(k) & 0%
\end{array}%
\right) ,~~g(k)=\left( 
\begin{array}{cc}
\xi_{k}-\Omega & -\Delta -i\alpha _{k} \\ 
\Delta+i\alpha_{k} & \xi _{k}+\Omega%
\end{array}%
\right) .
\end{equation}%
With the evolution of $k$ from $0$ to $2\pi $, the trajectory of the complex
function $z(k)\equiv\det g(k)$ forms a closed curve on the cylinder,
characterized by the following winding number 
\begin{eqnarray}
\gamma_+=\frac{1}{2\pi i}\oint_0^{2\pi}\frac{d z(k)}{z(k)}\,.  \label{z2}
\end{eqnarray}

If we take into account both mirror subspaces, the $\mathbb{Z}_{2}$
topological invariant can be formulated as $\nu=\gamma_+\mod 2$, consistent
with a previous theory~\cite{tritsc2}. Note that the TR symmetry requires the
winding numbers to be opposite for the two mirror subspaces: $%
\gamma_+=-\gamma_-$; by defining $\gamma_m=(\gamma_+-\gamma_-)/2$ we obtain $%
\nu=\gamma_m\mod 2$.

\begin{figure}[!h]
\includegraphics[width=3.95in]{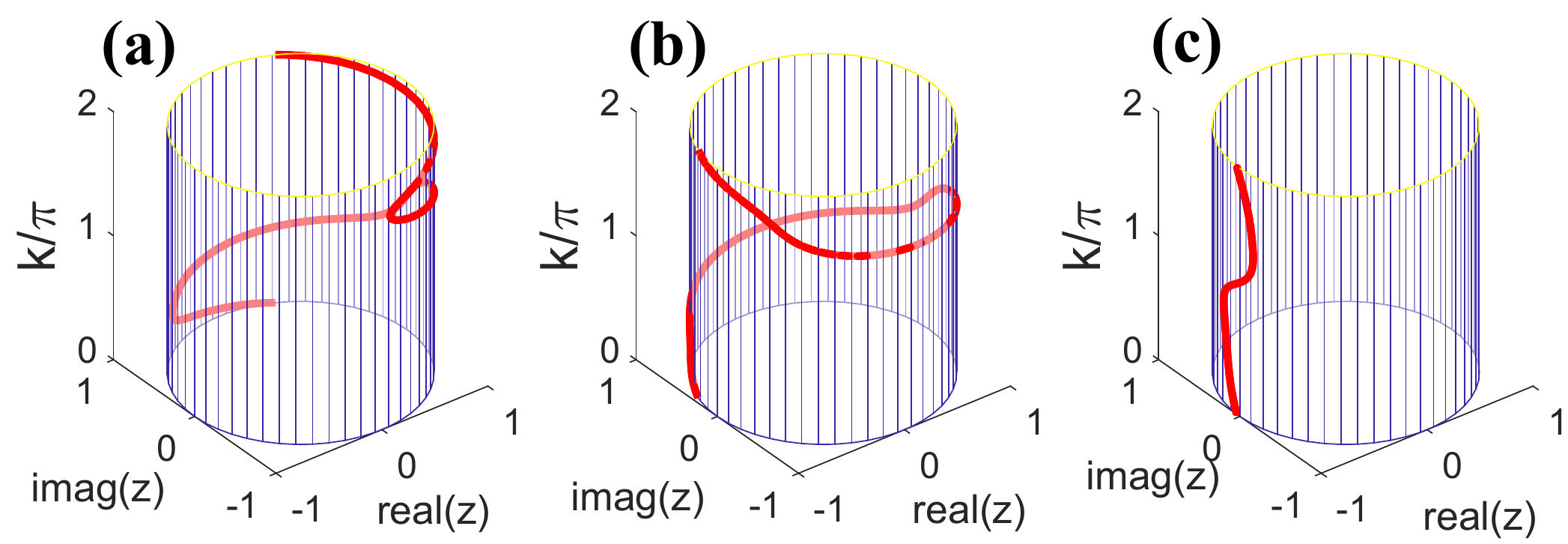}
\caption{Winding of ${z(k)}/{|z(k)|}$ in different phases. (a)~Normal
superfluid, $\Omega=0.5$. (b)~TR invariant topological superfluid, $\Omega=2$%
. (c)~Normal superfluid, $\Omega=5$. We have used $\protect\mu=-2$, $%
\Delta=1 $, $t=1$, $\protect\alpha=1$, and $t_{\perp}=0.5$; the phase
boundaries [see Eq.~(6) in the main text] are $\Omega_{c_1}\approx1.002$ and 
$\Omega_{c_2}\approx4.0615$.}
\label{fig_winding}
\end{figure}

An intuitive view of the trajectories of $z(k)$ in different phases is
illustrated in Fig.~\ref{fig_winding}. The system is a normal superfluid
when $|\Omega|<\Omega _{c_{1}}$. The evolution of $k$ from $0$ to $2\pi$
results in a contractible path on the cylinder surface [Fig.~\ref%
{fig_winding}(a)], indicating $\nu=\gamma_m=0$. For the TR invariant
topological (mirror) superfluid with $\Omega_{c_{1}}<|\Omega|<\Omega
_{c_{2}} $, the path is non-contractible around the cylinder [Fig.~\ref%
{fig_winding}(b)], indicating $\nu=\gamma_m=1$. Further increasing Zeeman
field to $|\Omega|>\Omega_{c_{2}}$, the system reenters into the normal
superfluid phase with a contractible path [Fig.~\ref{fig_winding}(c)]. The
validity of the $\mathbb{Z}_2$ invariant $\nu$ has been further confirmed by
the quasiparticle spectrum under the open boundary condition. For
topological superfluid, there exists a pair of degenerate zero modes, i.e.,
Majorana Kramers pair (MKP), localized on each end due to the TR symmetry.
For normal superfluid, no zero modes emerge in the bulk gap.

\subsection{II. 2D Case}

For the 2D Hamiltonian Eq.~(8) in the main text, the TR symmetry is $%
\mathcal{\tilde{T}}\mathcal{H}^{BdG}(k_{x},k_{y})\mathcal{\tilde{T}}^{-1}=%
\mathcal{H}^{BdG}(-k_{x},k_{y})$, which relates $(k_{x},k_{y})$ and $%
(-k_{x},k_{y})$. For each $k_{y}$, the effective 1D system belongs to the
symmetry class DIII, which is characterized by a $\mathbb{Z}_{2}$ invariant $%
\nu _{k_{y}}$. As a consequence, there exist three topologically distinct
superfluid phases: 
\begin{equation}
\left\{ 
\begin{array}{ll}
\text{Normal superfluid},~~~~~~~~~~~~~~~~~~~~~~~~~~~~v_{k_{y}}=0~\text{for
all}~k_{y}; &  \\ 
\text{TR-invariant topological superfluid},~~~~~~v_{k_{y}}=1~\text{for all}%
~k_{y}; &  \\ 
\text{Dirac-nodal superfluid},~~~~~~~~~~~~~~~~~~~~~~~v_{k_{y}}=0~\text{or}~1.
& 
\end{array}%
\right.
\end{equation}%
For a TR invariant topological superfluid, there emerge MKP edge flat bands
from $k_{y}=-\pi $ to $\pi $ in the open boundary condition. The Dirac-nodal
superfluid is gapless, whereas the 1D Brillouin zones at fixed $k_{y}$
values are divided into several topologically distinct regions, labeled by
either $\nu _{k_{y}}=0$ or $\nu _{k_{y}}=1$. The bulk gap must close at the
Dirac points, and these Dirac points can be regarded as topological phase
transition points for the effective 1D models. Thus, in the open boundary
condition, there are MKP edge flat bands connecting the projected Dirac
points. Totally, there are four Dirac points as dictated by both the TR and
mirror symmetries.

\subsection{Validity of the double-chain Model}

Our primary model Eq.~(1) in the main text describes two coupled 1D Fermi
gases experiencing the same spin-orbit coupling (SOC) but opposite Zeeman
fields introduced by the Hermite-Gaussian beam. To generate the double-chain
structure, a double well trapping potential along the $y$ direction is
needed. The tunneling $t_{\perp }$ between the two Fermi gases (i.e., the
kinetic energy along the $y$ direction) depends on the depth of the double
well trapping potential in the $y$ direction, which is tunable. In the
following, we show the validity of the tight-binding model Eq.~(1) in the
main text and provide the details of our calculations and estimations. Since
the realized SOC $\alpha k_{x}\sigma _{y}$ is intrinsically 1D, the
movements along the $x$ and $y$ directions are independent. The dynamics
along the $y$ direction is governed by the following single-particle
Hamiltonian: 
\begin{equation}
H_{y}=-\frac{\hbar ^{2}\partial _{y}^{2}}{2m}+\Omega (y)\sigma _{z}+V_{%
\mathrm{trap}}(y).
\end{equation}%
Here $\Omega (y)=\Omega _{0}ye^{\frac{-y^{2}}{w^{2}}}$ is the effective
Zeeman field induced by the Hermite-Gaussian beam. The trapping potential
along the \textit{y} direction can be approximately described by $V_{\mathrm{%
trap}}(y)=\frac{1}{2}m\omega _{y}^{2}y^{2}+V_{0}\cos ^{2}(k_{0}y)$, although
in practice a double well optical lattice can be used. We numerically solve
the eigenstates of $H_{y}$, with the lowest two-fold degenerate eigenstates
shown in Fig.~\ref{tbwf}.

\begin{figure}[h]
\centering
\includegraphics[width=2.8in]{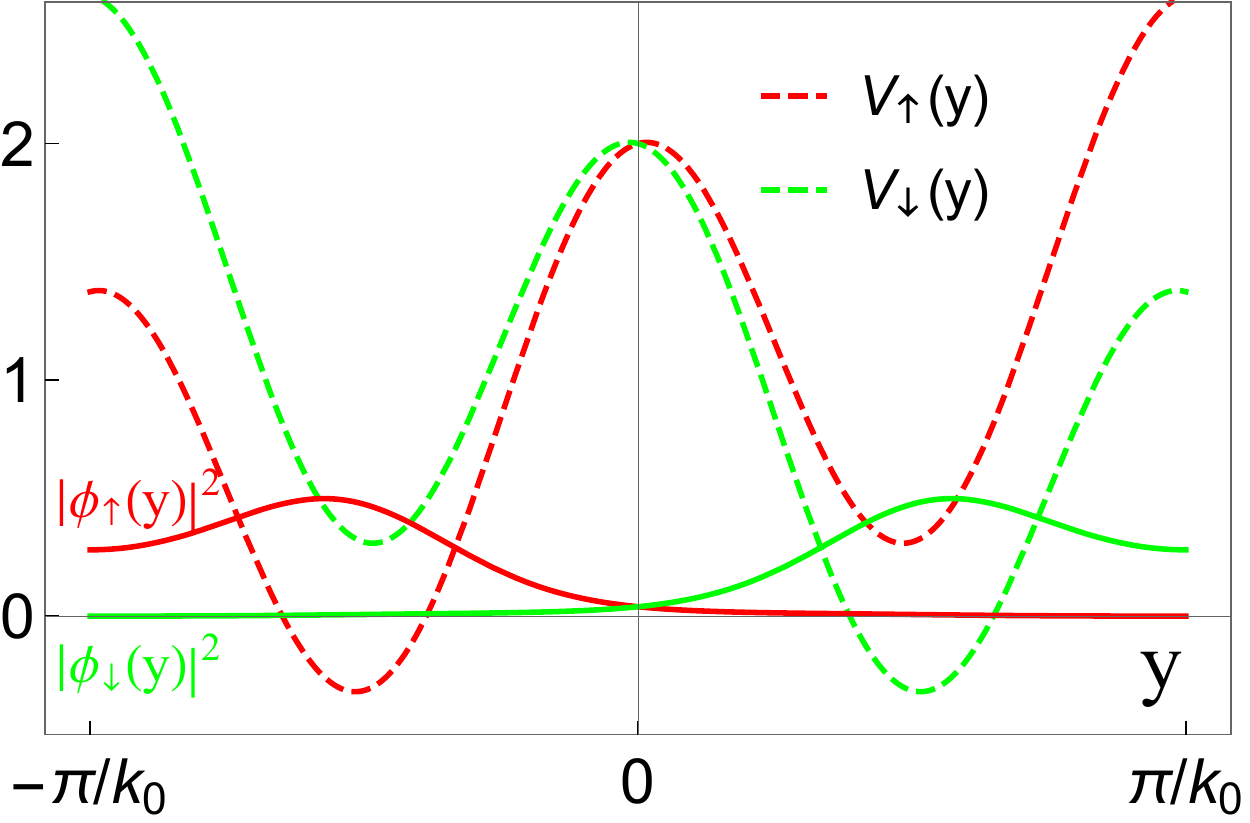}
\caption{Effective potentials $V_{\protect\sigma }(y)=\Omega (y)\protect%
\sigma _{z}+V_{\mathrm{trap}}(y)$ in units of $E_{R}$ and ground state
probabilities for spin-up and spin-down atoms. Typical experimental
parameters of $^{40}$K atoms are used. $k_{0}=2\protect\pi /\protect\lambda $
with $\protect\lambda =680$ nm, $E_{R}=2\protect\pi \hbar \times 10.8$ kHz.
For the harmonic trap, $\hbar \protect\omega _{y}=0.02E_{R}$; for the dipole
trap, $V_{0}=2E_{R}$. $\Omega _{0}=0.2k_{0}E_{R}$. The waist of the
Hermite-Gaussian beam is $2\protect\pi w=100\protect\lambda $.}
\label{tbwf}
\end{figure}

Obviously, the ground state mainly resides on the potential minimum for each
spin species, validating our tight-binding approximation in the $y$
direction. A rough approximation of $t_{\perp }$ can be obtained from the 1D
Mathieu equation $t_{\perp }\approx \frac{4}{\sqrt{\pi }}(\frac{V_{0}}{E_{r}}%
)^{3/4}e^{-2(\frac{V_{0}}{E_{R}})^{1/2}}=0.224E_{R}$. The Zeeman field $%
\Omega (y)$ can be controlled by tuning the Hermite-Gaussian beam. Note that
in our simulation the condition of large $\Omega (y)$ is not needed for the
tight-binding approximation, although a large Zeeman field indeed tends to
localize the two ground states at the potential minima.

\subsection{Continuum model}

For the dynamics along the $x$ direction (with 1D SOC), we have considered a
lattice model as described by Eq.~(1) in the main text. This lattice
regularization does not change the essential results. In this section, we
consider the continuum version of the 1D BdG Hamiltonian and demonstrate the
phase diagram and the emergence of MKPs in the TR invariant topological
(mirror) superfluid. As we shall see, the essential physics of the continuum
model is the same as that in the lattice model (as shown in the main text).

\subsection{I. Phase diagram}

In the same Nambu basis, the spin-orbit coupled double Fermi gases can be
described by the following continuum model Hamiltonian 
\begin{equation}
\mathcal{H}_{\mathrm{continuum}}^{\mathrm{BdG}}(k)={\large \left[\frac{\hbar
^{2}k^{2}}{2m}-\tilde{\mu}+2\alpha k\sigma _{y}-t_{\perp }s_{x}\right]\tau
_{z}+\Omega \sigma _{z}s_{z}+\tilde{\Delta}\tau _{x}.}  \label{contbdg}
\end{equation}%
Here $k$ is the particle momentum along ${x}$ direction, $\tilde{\mu}$ is
the chemical potential, and $\alpha ={\hbar ^{2}k_{r}}/{2m}$ is the SOC
strength, with $k_{r}$ the recoil momentum \cite{1dsoc6,1dsoc7}. The lattice
model Eq.~(1) in the main text can be obtained from the above continuum
model via the standard substitution: $\sin k\leftrightarrow k$, $\cos
k\leftrightarrow 1-{k^{2}}/{2}$. The pairing order parameter is defined as $%
\tilde{\Delta}=g\sum_{k}\langle c_{-k\downarrow }c_{k\uparrow }\rangle $,
with $g$ ($<0$) the attractive inter-particle interaction. The transverse
tunneling $t_{\perp }$ can be tuned by adjusting the trap depth in the ${y}$
direction. For simplicity, we take the recoil energy $E_{r}={\hbar
^{2}k_{r}^{2}}/{2m}$ and the recoil momentum $k_{r}$ as natural energy and
momentum units (or equivalently by setting $\hbar =2m=1$ in Eq. (\ref%
{contbdg})).

For the above continuum model Hamiltonian, the transition between the
topological and normal superfluids (i.e., the condition for gap closure) is
determined by the critical Zeeman field strength 
\begin{equation}
\Omega _{c}=\sqrt{(\frac{t_{\perp }^{2}}{4}-\tilde{\mu})^{2}+\tilde{\Delta}%
^{2}}\,.  \label{bcc}
\end{equation}%
For $\Omega <\Omega _{c}$ the superfluid is normal, whereas for $\Omega
>\Omega _{c}$, the superfluid is topological and hosts MKPs on its boundary.
Now we self-consistently calculate the phase diagram of the continuum model.
At zero temperature, the thermodynamical potential $\Omega _{\mathrm{TP}}=-({%
1}/{\beta})\log \text{Tr}e^{-\beta \lbrack \sum_{k}H_{\mathrm{continuum}}^{%
\mathrm{BdG}}(k)-\tilde{\mu}N]}$ (with $\beta ={1}/{k_{B}T}$) reduces to 
\begin{equation}
\Omega _{\mathrm{TP}}=\sum_{k}(k^{2}-\tilde{\mu})+\sum_{E_{n,k}<0}E_{n,k}-%
\frac{|\tilde{\Delta}|^{2}}{g}.
\end{equation}%
Here $E_{n,k}$ is the $n$-th eigenenergy of $H_{\mathrm{continuum}}^{\mathrm{%
BdG}}(k)$. The phase diagram can be analytically obtained by minimizing $%
\Omega _{\mathrm{TP}}$ with respect to the order parameter $\tilde{\Delta}$,
i.e., ${\partial \Omega _{\mathrm{TP}}}/{\partial \tilde{\Delta}}=0$.
However, due to its complexity, we extract the phase diagram numerically.
Note that we should take $\tilde{\mu}=\mu +2$ to compare with the phase
diagram of the lattice model in the main text. (For $k=0$, the kinetic
energy is $-2$ in the lattice model yet $0$ in the continuum model.)

\begin{figure}[h]
\includegraphics[width=4.5in]{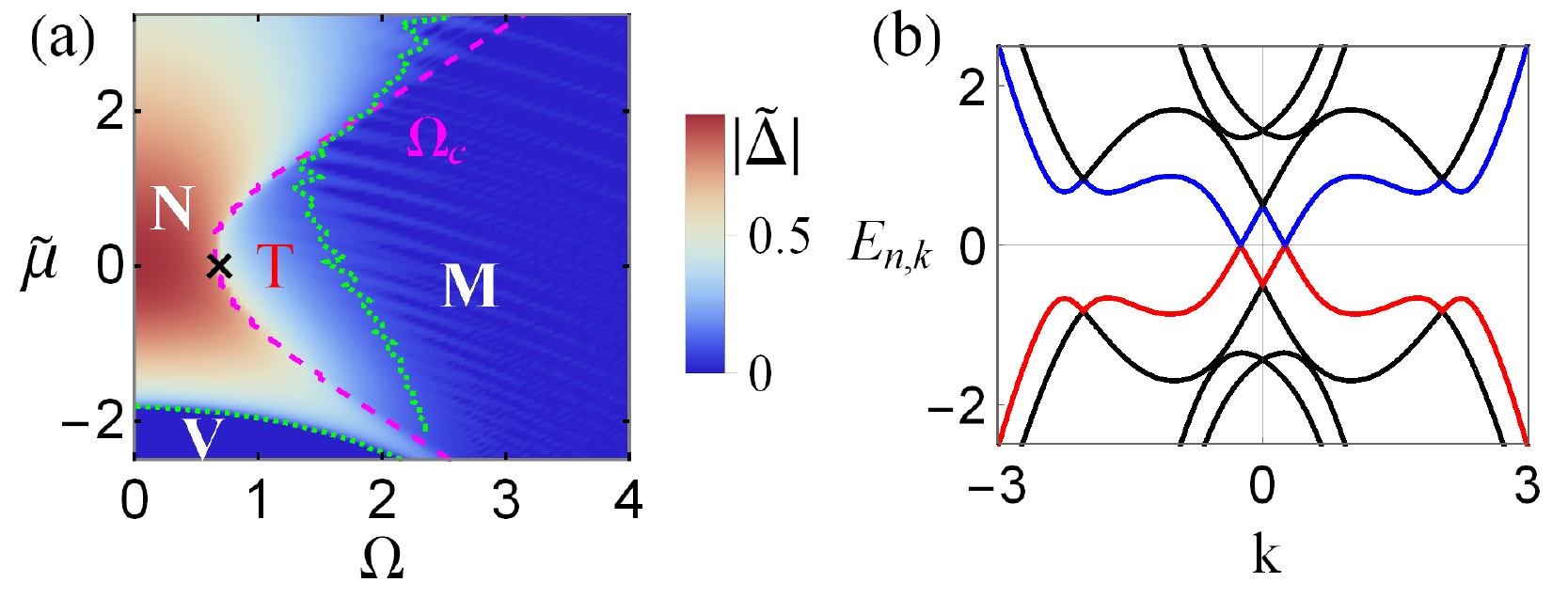}
\caption{(a)~Phase diagram of the continuum model. The dashed magenta line
denotes the critical Zeeman field $\Omega _{c}$ determined by Eq.~(\protect
\ref{bcc}). The dotted cyan lines mark the phase boundaries with reference
pairing $|\tilde{\Delta}|=0.05$. Totally there are four different phases:
vacuum phase (V), normal superfluid (N), topological superfluid (T) and
metal phase (M). (b)~Band crossings at the phase transition point shown by
\textquotedblleft $\times $\textquotedblright\ in (a); $\tilde{\protect\mu}%
=0 $, $\Omega \approx 0.68$, and $t_{\perp }=0.5$.}
\label{phase_diag_conti}
\end{figure}

The phase diagram in $\Omega -\tilde{\mu}$ plane is illustrated in Fig.~\ref%
{phase_diag_conti}(a), which is quite similar to that of the lattice model
[see Fig.~2(a) in the main text], especially in the low-energy regime (i.e.,
small $\tilde{\mu}$). Besides the vacuum phase (V) and metal phase (M) with
vanishing pairing, there are two types of superfluid phases with finite
pairing order parameters $\tilde{\Delta}$. The topological superfluid (T)
resides between the normal superfluid (N) and metal phase, with the critical
Zeeman field strength $\Omega _{c}$ determined by Eq.~(\ref{bcc}). Note
that, due to the lack of lattice, there is no band insulator phase. The
quasiparticle spectrum at the phase transition point between the normal and
topological superfluids is shown in Fig.~\ref{phase_diag_conti}(b), where
the central two bands cross at $k=\pm {t_{\perp }}/{2}$.

\subsection{II. Emergence of MKPs on the boundary}

The topological properties of the above TR-invariant topological superfluid
phase are characterized by the appearance of MKPs on the boundary. This can
be easily demonstrated by self-consistently solving the following real-space
BdG equation: 
\begin{equation}
H_{\mathrm{BdG}}(x)\Psi _{n}(x)=E_{n}\Psi _{n}(x),  \label{rbdg}
\end{equation}%
where $\Psi _{n}(x)\equiv \lbrack u_{n}^{1\uparrow }(x),u_{n}^{1\downarrow
}(x),u_{n}^{2\uparrow }(x),u_{n}^{2\downarrow }(x),v_{n}^{1\uparrow
}(x),v_{n}^{1\downarrow }(x),v_{n}^{2\uparrow }(x),v_{n}^{2\downarrow
}(x)]^{T}$ is the eight-component wave function of the real-space
Hamiltonian $H_{\mathrm{BdG}}(x)$ obtained by the substitution $k\rightarrow
-i\partial _{x}$ in the momentum-space Hamiltonian Eq.~(\ref{contbdg}). The
local pairing of each Fermi gas ($s=1,2$) is represented by $\tilde{\Delta}%
^{s}(x)=g\sum_{n}[u_{n}^{s\downarrow }v_{n}^{s\uparrow \ast
}f(-E_{n})+u_{n}^{s\uparrow }v_{n}^{s\downarrow\ast }f(E_{n})]$. Here $%
f(x)=1/(e^{x/k_{B}T}+1)$ the Fermi distribution function, and at zero
temperature it reduces to the Heaviside step function.

To examine the existence of MKPs in the topological superfluid, we employ an
open boundary condition at $x=0$ and $x=L$. The wave function can be
expanded by a set of base functions as follows: 
\begin{eqnarray}
u^{s\uparrow}_{n}=\sum_m A_{nm}^{s\uparrow}\sqrt{\frac{2}{L}}\sin(\frac{m\pi
x}{L}),~~u^{s\downarrow}_{n}=\sum_m A_{nm}^{s\downarrow}\sqrt{\frac{2}{L}}%
\sin(\frac{m\pi x}{L}),  \notag \\
v^{s\uparrow}_{n}=\sum_m B_{nm}^{s\uparrow}\sqrt{\frac{2}{L}}\sin(\frac{m\pi
x}{L}),~~v^{s\downarrow}_{n}=\sum_m B_{nm}^{s\downarrow}\sqrt{\frac{2}{L}}%
\sin(\frac{m\pi x}{L}).
\end{eqnarray}
To obtain the eigenspectrum of Eq.~(\ref{rbdg}), we diagonalize the
Hamiltonian (of size $8N_c\times 8N_c$) in the chosen basis. In our
calculations, the truncation number of the basis is $N_c=100$.

\begin{figure}[!h]
\includegraphics[width=4.5in]{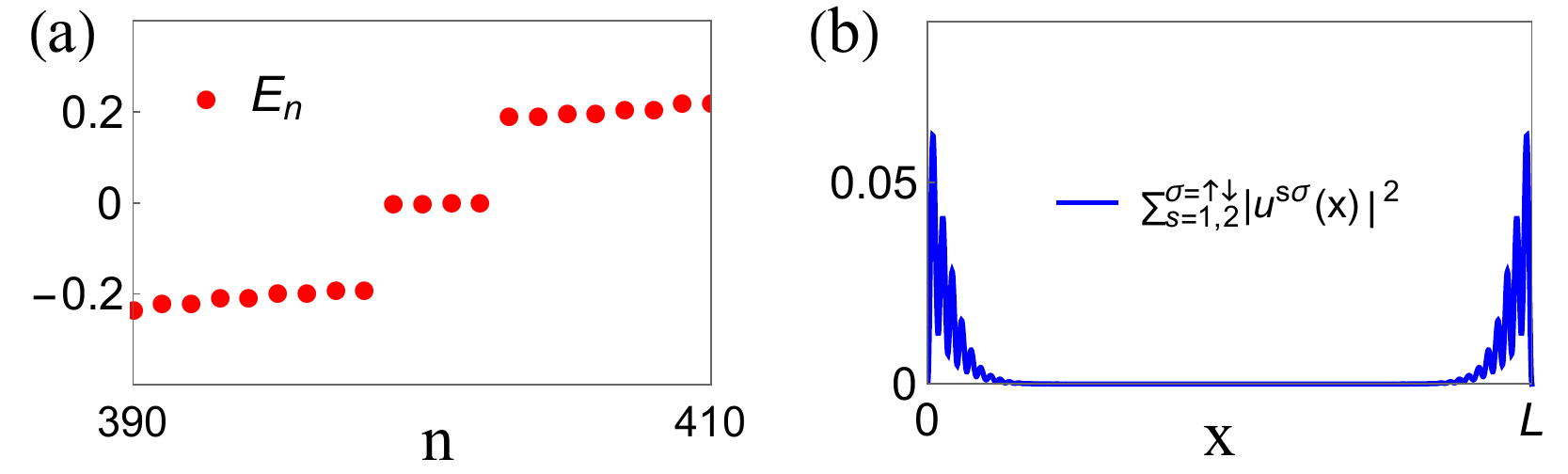}
\caption{(a) Quasiparticle spectrum in the open boundary condition, with
four-fold degenerate zero modes. (b) Spatial distribution of wave function $%
\Psi_n(x)$ with $n=4N_c+1$. Here $t_{\perp}=0.5$, $\tilde{\protect\mu}=0$, $%
\Omega=1$, and $N_c=100$.}
\label{wf_conti}
\end{figure}

The quasiparticle spectrum in the open boundary condition is plotted in Fig.~%
\ref{wf_conti}(a). There exist four-fold degenerate zero modes in the bulk
pairing gap due to the TR and particle-hole symmetries and the existence of
two ends. Take the $n=4N_{c}+1$ state as an example, the wave function is
mainly localized at the two ends ($x=0$ and $L$), as seen in Fig.~\ref%
{wf_conti}(b). By contrast, for a normal superfluid, the quasiparticle
spectrum is fully gapped without any in-gap zero mode. All these results
clearly show that the essential physics is much the same for the lattice
model considered in the main text and the continuum model discussed here.

\subsection{Digital micromirror device and application}

In this section, we explain how to obtain the desired laser fields using the
digital micromirror device (DMD). The basic principle is sketched in Fig.~%
\ref{dmd}.

\begin{figure}[h]
\centering
\includegraphics[width=3.5in]{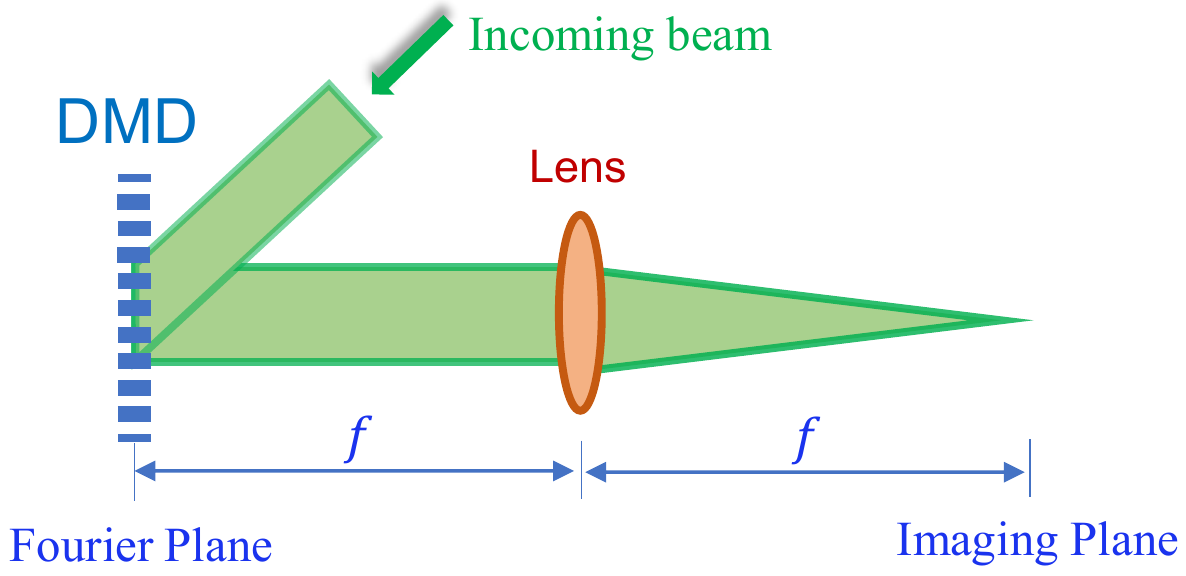}
\caption{Schematics of holographic shaping of laser beams.}
\label{dmd}
\end{figure}

DMD is an optomechanical spatial light modulator, consisting of many square
mirrors. Each mirror can be individually switched between two tilt states ($%
+12$ ``on'' and $-12$ ``off'' orientations) and acts as a basic diffractive
element under coherent illumination due to its small size. The 2D mirror
array forms a reflective grating. The DMDs are employed holographically in
the Fourier plane of an imaging system, which enables both local amplitude
and phase control simultaneously despite that the direct DMD modulations is
binary in intensity. The incoming light is not only reflected into one of
the two directions, but it also has a number of diffraction orders. For the $%
m$-th diffraction order (in the $y$ direction), the outgoing light field of
a single slit of width $a$ centered at $y_0$ is 
\begin{equation}
E_{\mathrm{out}}=E_{\mathrm{in}}\frac{\sin(\pi m a)}{\pi m} e^{i 2\pi m y_0}.
\end{equation}

It is clear that by shifting the slits relative to some reference point, the
phase of the diffracted wave fronts can be tuned, while the intensities of
the outgoing beams are determined by the width of the slits. Applying these
phase and amplitude modifications locally, arbitrary wave fronts can be
generated in this way. The downside of the above DMD setup is its relatively
low efficiency in laser power. This can be optimized by fulfilling the
so-called blazing condition and increasing the intensity of incoming beams.
Currently, the DMD-generated laser beams have been widely used to address
individual atoms in optical lattices and to manipulate their dynamics.

\subsection{Experimental detection of MKPs}

In this section, we discuss how to detect the MKPs in the topological
superfluid phase using the spatially resolved radio-frequency spectroscopy~%
\cite{rf1,rf2,rf3,ldosrf1,ldosrf2}. The radio-frequency (rf) field is a
probe field to induce single-particle excitation from the fermionic state to
an unoccupied fluorescent probe state. 
The rf signal is directly related to the local density of states (LDOS),
similar to that of scanning tunneling microscope. The LDOS for the chain-$s$
contains both spin components $\rho_{s}(j,\omega)=\rho_{s\uparrow}(i,%
\omega)+\rho_{s\downarrow}(i,\omega)$. In the BdG representation, 
\begin{eqnarray}
\rho_{s\sigma}(j,\omega)=\frac{1}{2}\sum_{\eta}|u^{s\sigma}_{\eta}(j)|^2%
\delta(\omega-E_{\eta})+|v^{s\sigma}_{\eta}(j)|^2\delta(\omega+E_{\eta}).
\end{eqnarray}

We numerically calculate the LDOS for the double-chain lattice system
(Eq.~(1) in the main text) under a weak harmonic trapping potential $%
V(j)=w_{x}^{2}(j-\frac{L+1}{2})^{2}$. For $w_{x}\ll 1$, i.e., the
characteristic length of the harmonic trap is march larger than other length
scales, local density approximation (LDA) $\mu (j)=\mu -V(j)$ can be used to
analyze the local excitations in real space.

The main results are shown in Fig.~\ref{dos}. The harmonic trap separates
the system into topologically distinct regions: TR-invariant topological
superfluid around the trap center and normal superfluids at the two wings.
The boundary between two spatially separated phases is approximately
determined by substituting $\mu \rightarrow \mu (j)$ in Eq.~(6) in the main
text. According to bulk-boundary correspondence, a MKP emerges at each
topological boundary. Note that due to the finite length of the system, the
MKPs at the two boundaries slightly hybridize and form quasiparticle levels
(two-fold degenerate) and quasihole levels (two-fold degenerate). We denote
the wave functions of the degenerate quasiparticle levels as $\phi
_{a}=(u_{a}^{1\uparrow },u_{a}^{1\downarrow },u_{a}^{2\uparrow
},u_{a}^{2\downarrow },v_{a}^{1\uparrow },v_{a}^{1\downarrow
},v_{a}^{2\uparrow },v_{a}^{2\downarrow })$ ($a=\text{I},\text{II}$). Fig.~%
\ref{dos}(a) plots their spatial distributions, which are localized on the
boundaries between the topological superfluid and normal superfluids.
Furthermore, as verified by our numerics, the two-fold degenerate
quasiparticle states are related by the TR symmetry as: $\phi _{\text{II}}=%
\mathcal{\tilde{T}}\phi _{\text{I}}$ and $\phi _{\text{I}}=-\mathcal{\tilde{T%
}}\phi _{\text{II}}$. In the form of their components, $u_{\text{I}%
}^{1\uparrow }=u_{\text{II}}^{2\downarrow }$, $u_{\text{I}}^{1\downarrow
}=-u_{\text{II}}^{2\uparrow }$, $u_{\text{I}}^{2\uparrow }=u_{\text{II}%
}^{1\downarrow }$, and $u_{\text{I}}^{2\downarrow }=-u_{\text{II}%
}^{1\uparrow }$. Note that the wave functions can always be chosen as real
since the real-space BdG Hamiltonian is real.

\begin{figure}[h]
\includegraphics[width=6in]{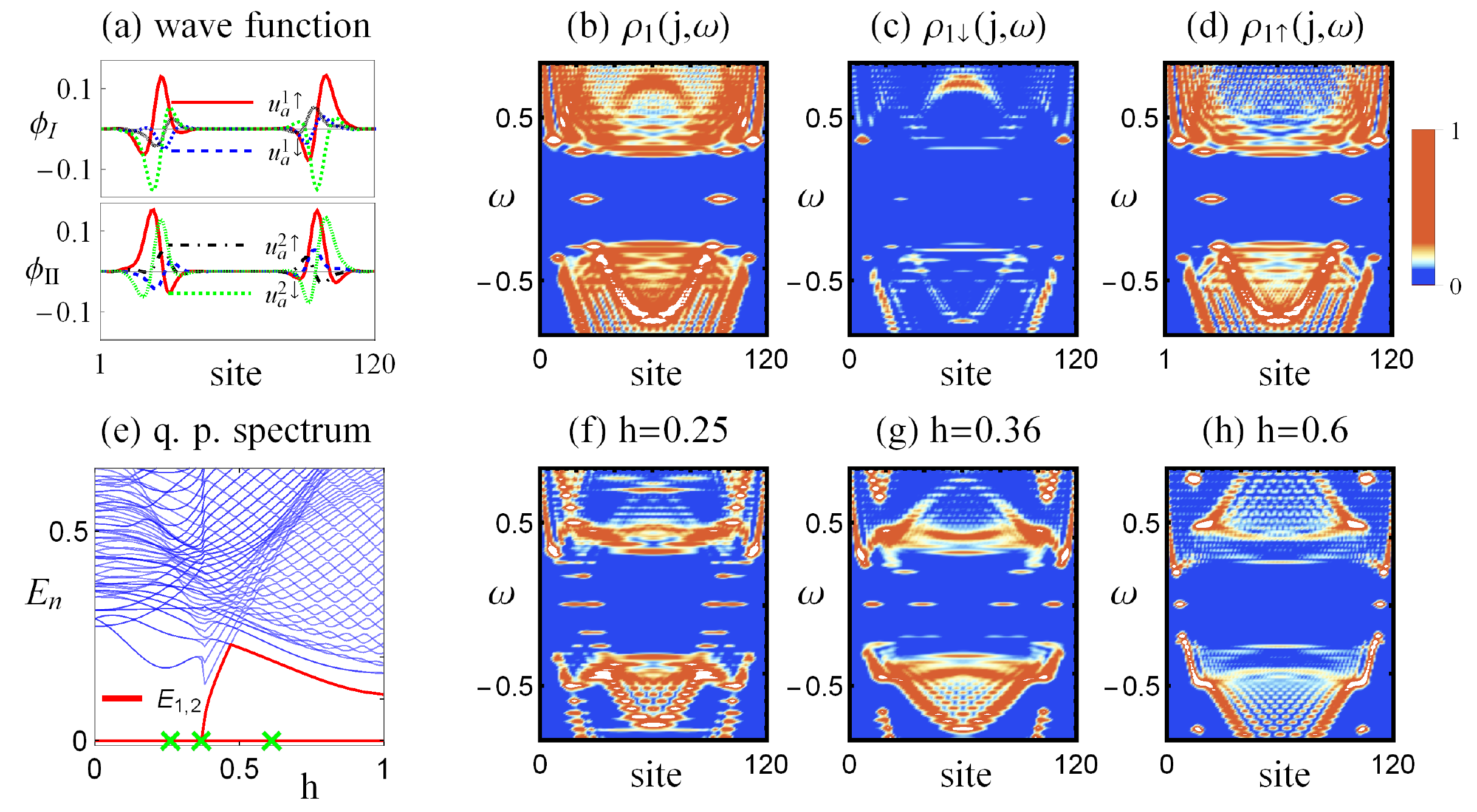}
\caption{(a) Wave functions of the MKP related by TR symmetry. (b) LDOS $%
\protect\rho_1(j,\protect\omega)$ for chain-1. (c) Spin-resolved LDOS $%
\protect\rho_{1\uparrow}(j,\protect\omega)$. (d) Spin-resolved LDOS $\protect%
\rho_{1\downarrow}(j,\protect\omega)$. For (a)-(d), $\Omega=1$. (e)
Quasiparticle spectrum versus TR breaking Zeeman field $h$. (f)-(h) LDOS $%
\sum_{s=1}^{2}\protect\rho_{s}(j,\protect\omega)$ for three typical $h$
values, as labeled by ``cross'' in (e). The harmonic trapping potential is $%
w_x=0.025$, and other parameter values are the same as those used in
Fig.~2(c) in the main text: $\protect\alpha=1$, $t_{\perp}=0.5$, and $t=1$
(set as the energy unit).}
\label{dos}
\end{figure}

The above (hybridized) MKPs can be read out directly from the LDOS (note
that $\rho_{1}(j,\omega)=\rho_{2}(j,\omega)$ because of the TR symmetry), as
illustrated in Fig.~\ref{dos}(b). These zero-energy states localized at the
topological boundaries are protected by the pairing gap. For the chain-1,
the spin-resolved LDOS is shown in Figs.~\ref{dos}(c)-(d). (For the chain-2, 
$\rho_{2\uparrow}(j,\omega)=\rho_{1\downarrow}(j,\omega)$ and $%
\rho_{2\downarrow}(j,\omega)=\rho_{1\uparrow}(j,\omega)$). Our numerics show
that these zero-energy states are mainly composed of the spin-up component
for the chain-1 and spin-down for the chain-2, indicating the \textit{%
emergent effective} TR symmetry of the system.

Further, the above LDOS signatures of MKPs are intrinsically different from
that of a single Majorana mode. To reveal its double degeneracy, we note
that the existence of MKPs are protected by TR symmetry, which would be
broken by adding a small Zeeman field $h\sigma _{z}$ (equal for both
chains). In our experimental scheme, the TR-invariant staggered Zeeman field 
$\Omega s_{z}\sigma _{z}$ is generated by the Hermite-Gaussian beam. The
TR-breaking Zeeman field can be easily induced by shifting the system along
the ${y}$ direction. The quasiparticle spectrum is shown in Fig.~\ref{dos}%
(e). For small $h$, the lowest quasiparticle excitation is still two-fold
degenerate, however the boundary Majorana pair start to separate spatially:
one of them moves toward the trap center while the other one moves toward a
trap wing slowly [Fig.~\ref{dos}(f)]. At a critical strength $h\approx 0.36$%
, two Majorana modes coming from the MKPs of different sides collide in the
trap center [Fig.~\ref{dos}(g)] and annihilate each other, leaving only one
Majorana mode on each wing with further increasing $h$ [Fig.~\ref{dos}(h)].

The above spatial separation of the MKPs in a harmonic trap can be analyzed
using the LDA. From the bulk-edge correspondence, the Majorana zero modes
should emerge at the boundary between topologically distinct 
regions, which is determined by the gap closure condition. We start from
the uniform system. With an applied Zeeman field (the total Hamiltonian is $%
\mathcal{H}_{k}^{\mathrm{BdG}}+h\sigma _{z}$), the gap closure conditions
are given by ($\xi _{k}=-2t\cos k-\mu $): 
\begin{eqnarray}
\xi _{k}^{2}+\Delta ^{2}+(t_{\perp }^{2}-4\alpha ^{2}\sin k^{2})-h^{2}
&=&\Omega ^{2},  \label{cond1} \\
\frac{4\alpha ^{2}\sin ^{2}k-\Delta ^{2}+\sqrt{(4\alpha ^{2}\sin
^{2}k+\Delta ^{2})^{2}+4\xi _{k}^{2}h^{2}}}{2}+h^{2} &=&t_{\perp }^{2}.
\label{cond2}
\end{eqnarray}%
Obviously, in the limit of $h\rightarrow 0$, the above conditions reduce to
the TR invariant case (see Eq.~(5) in the main text). We locally substitute $\mu$ 
by $\mu (j)=\mu -V(j)$ in the above two equations for the harmonic trap
case, from which the positions of the zero modes can be determined. In
Fig.~\ref{mkpsplit}, we plot the positions of the split Majorana modes as a
function of $h$ by numerically solving the gap closure conditions. At $h=0$,
there is one MKP distributed at each side symmetrically. With increasing $h$%
,  we can find two solutions for each side, corresponding to two split
Majorana modes from the same MKP. The numerical solutions agree well with
the LDOS pictures shown in Fig.~\ref{dos}(f-h). The splitting becomes faster
at larger $h$, and finally at the critical Zeeman field only one solution
survives, consistent with the annihilation of two Majorana modes at the
trap center.

\begin{figure}[h]
\includegraphics[width=4in]{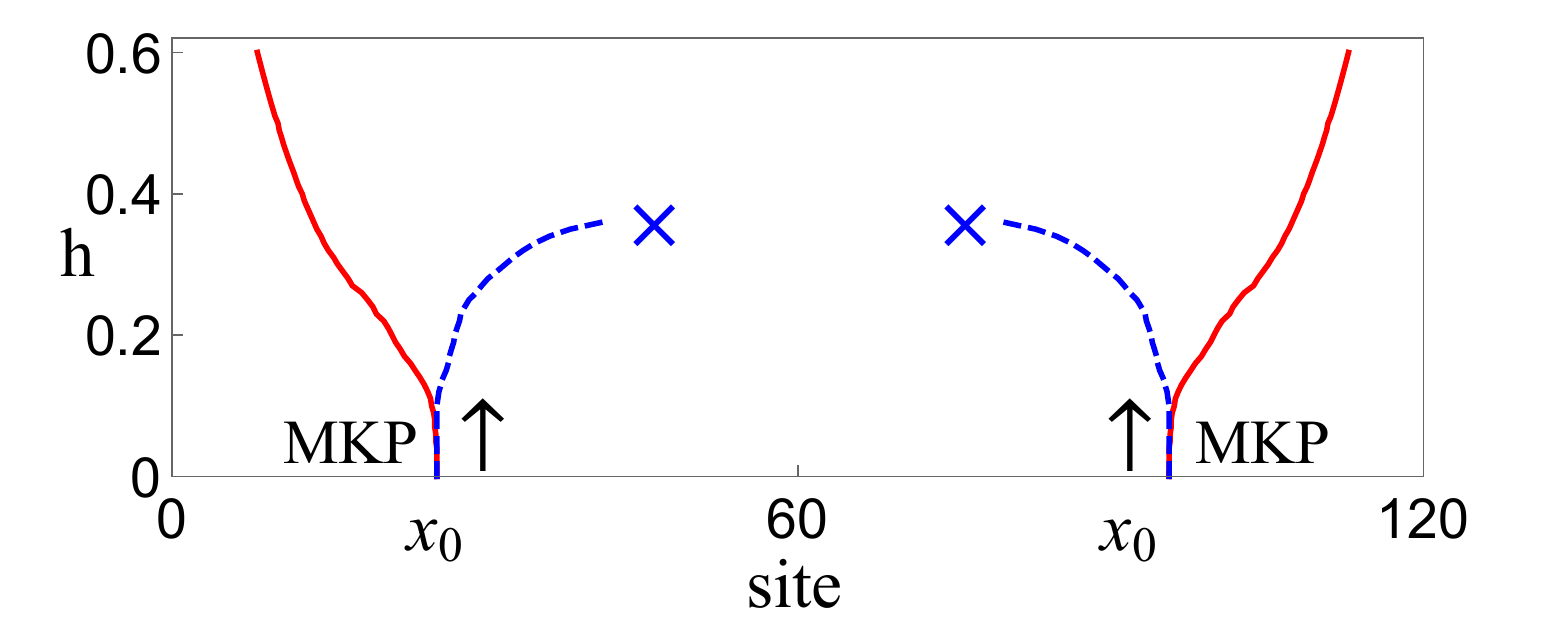}
\caption{Spatial splitting of a MKP into two Majorana zero modes in a harmonic
trap under a Zeeman field $h$. The positions of these zero modes (red lines
and blue dotted lines) are determined by Eq.~(\protect\ref{cond1}) and~(\protect
\ref{cond2}) with the parameters from the self-consistent BdG calculations. At $h=0$%
, there is one MKP at each side. $x_0$ denotes the original position of a
MKP.}
\label{mkpsplit}
\end{figure}

It is worth to mention that, although in general it is hard to write the
dependence of $\mu $ on $h$ in an explicit form analytically, we can analyze
the splitting of the MKP for a small $h$. Denote the original position of
the MKP in a harmonic trap as $x_{0}$. From Eq.~(\ref{cond2}), we have $%
4\alpha ^{2}\sin ^{2}k=t_{\perp }^{2}-(\xi _{k}^{2}+1)h^{2}+o(h^{2})$.
Substituting this into Eq.~(\ref{cond1}), we get $\xi _{k}^{2}=(\Omega
^{2}-\Delta ^{2})(1-h^{2})+o(h^{2})$, yielding the following relation for
the chemical potential: 
\begin{equation}
\mu (h)=-2t\cos k-\xi _{k}=\mu \pm \frac{1}{2}[\xi _{k,0}+\frac{t(\xi
_{k,0}^{2}+1)}{2\alpha ^{2}}]h^{2}+o(h^{2}).  \label{chemical}
\end{equation}%
Here $\mu $ is the chemical potential without adding the Zeeman field, and $%
\xi _{k,0}^{2}=\Omega ^{2}-\Delta ^{2}$ (Eq.~(5) in the main text). Now
using the LDA and taking the derivative on both sides of Eq. ({\ref{chemical}}), we
can get the following relation:
\begin{equation}
\delta x=\pm \frac{\lbrack \xi _{k,0}+\frac{t(\xi _{k,0}^{2}+1)}{2\alpha ^{2}%
}]h^{2}}{4\omega _{x}^{2}x_{0}},
\end{equation}%
which determines the shifting of the spatial phase boundary by applying a small
Zeeman field. Here the $+$ and $-$ signs correspond to the right- and left-moving zero modes, respectively. 
The above analysis clearly shows $\delta x\sim h^{2}$, 
explaining the slow initial splitting rate of MKP. This point has
also been verified directly from our LDOS results. 
\end{document}